\documentclass[floatfix,aps,twocolumn,prb,superscriptaddress]{revtex4-1}
\usepackage{graphicx}
\usepackage{dcolumn}
\usepackage{bm}
\usepackage[T1]{fontenc}
\usepackage[latin9]{inputenc}
\usepackage{textcomp}
\usepackage{amsmath}
\usepackage{mathrsfs}
\usepackage{graphicx}
\usepackage{amssymb}
\usepackage{units}
\usepackage{xcolor}

\begin{document}

\title{New perspectives on superfluidity in resonantly--driven polariton fluids}

\author{Ivan Amelio}
\affiliation{INO-CNR BEC Center and Dipartimento di Fisica, Universit{\`a} di Trento, 38123 Povo, Italy}
\author{Iacopo Carusotto}
\affiliation{INO-CNR BEC Center and Dipartimento di Fisica, Universit{\`a} di Trento, 38123 Povo, Italy}


\begin{abstract}
In this paper we discuss, within the Gross--Pitaevskii framework, superfluidity, soliton nucleation, and instabilities in a non-equilibrium polariton fluid injected by a spatially localized and continuous-wave coherent pump and flowing against a defect located outside the pump spot. In contrast to equilibrium condensates, the steady-state solutions of the driven-dissipative equations in this specific geometry hardly show a clean superfluid flow around the defect and rather feature a crossover from shallow to deep soliton-like perturbation. This is explained in terms of the properties of one-dimensional flows, in particular their weak dependence on the pump parameters and their rapid transition to a super-sonic regime under the effect of the quantum pressure; such a highly nonlinear behaviour calls for quantitative experimental tests of the underlying Gross--Pitaevskii equation. The role of disorder and of a incoherent reservoir in inducing non-stationary behaviours with moving vortices is also highlighted. 
\end{abstract}

\date{\today}

\maketitle


\section{Introduction}

Exciton--polaritons are mixed bosonic quasi-particles that combine the interesting properties of the underlying excitons and photons. From the former, they inherit strong binary interactions. Thanks to latter component, they can be easily injected and detected by optical means, which allows for accurate control and imaging of their dynamics.
The dynamics of polariton fluids has been extensively studied in a mean--field, hydrodynamic context based on the driven-dissipative Gross--Pitaevskii equation (GPE) \cite{carusotto2004}, but experimental and theoretical work has been so far guided by the analogies to equilibrium superfluids~\cite{carusotto2013}. In particular, Bose--Einstein condensation \cite{kasprzak2006}, superfluidity \cite{amo2009} and turbulence \cite{amo2011, sanvitto2011, nardin2011} have been experimentally reported in the last decade.


In this spirit, the microscopic solid-state physics of the semiconductor microcavity device has been so far considered as a limitation, but might become a useful asset in view of new developments. For instance, several recent experiments~\cite{stepanov2018, sarkar2010,walker2017} have pointed out marked deviations from the basic GPE model in the dispersion of the collective Bogoliubov excitations~\cite{stepanov2018}, in the bistable behaviours of spin--mixtures~\cite{sarkar2010}, and in the dynamics of solitons~\cite{walker2017}. All these phenomena provide evidence for the presence of an incoherent reservoir of excitations that spontaneously forms via polariton absorption even under a coherent resonant pump. In addition to its own interest for quantum fluids of light, polariton superfluidity then turns out to be a powerful tool to access the microscopic physics of the underlying material.

In this work, we report a theoretical study of the effect of driving and dissipation on the superfluidity properties of polariton fluids and of the consequences of the incoherent reservoir on the superfluid dynamics. While earlier work has focussed on spatially homogeneous fluids~\cite{carusotto2004} and has formulated generalized non-equilibrium versions of the Landau critical speed~\cite{wouters2010, amelio2019b}, we focus here on the effect of a finite pump spot and of its consequences on the density and velocity profile of the fluid. 
Striking observable features in the interaction of the fluid with a static defect located outside the pumped region are found and discussed: in contrast to the equilibrium case of, e.g. ultracold atomic gases, the steady-state solutions of the driven-dissipative equations for polariton fluids hardly show a clean superfluid flow around the defect, if in its location the driving field is strictly zero. 
The usual transitions from a superfluid flow to a turbulent regime to soliton nucleation~\cite{carusotto2013} appear to be replaced by a crossover from a shallow to a deep soliton-like perturbation. This new theoretical insight calls for a new generation of quantitative experimental studies of polariton superfluidity, which will shine new light on the general physics of non-equilibrium superfluidity as well as on the microscopic processes at play in the underlying semiconductor material.

%
%
%

The structure of the article is the following.
In Sec.~\ref{sec:Preliminary} we review the driven-dissipative GPE theory of resonantly driven polariton fluids. In sec.\ref{ssec:1D}, we discuss its solution in 1D geometries under a spatially localized pump. Acceleration of polaritons by the quantum-pressure-driven density gradient outside the pump spot is shown to quickly induce a transition from sub- to super-sonic flow in the close vicinity of the pump spot. In Sec.~\ref{sec:LinGeo}, we show how this feature restricts our possibility to explore the superfluid hydrodynamics in the undriven region outside the pump. The effect of disorder and of the incoherent reservoir on the stationarity and the coherence properties of the fluid is also unveiled. Additional effects are investigated in Sec.~\ref{sec:CircularSpot}, where the consequences of a long-tailed pump and of a circular geometry are explored.
Conclusions and future perspectives are finally discussed in  Sec.\ref{sec:conclu}.

\section{Driven-dissipative Gross-Pitaevskii equation}

\label{sec:Preliminary}

In this section, we review the driven-dissipative Gross-Pitaevskii equation that is currently used to describe the hydrodynamics of exciton--polaritons fluids under a coherent drive~\cite{carusotto2013}.

At the mean--field level, a polariton fluid driven by a monochromatic coherent field of spatial amplitude $F(\mathbf{r})$ and frequency $\omega_p$ can be described by a single--particle wavefunction $\psi(\mathbf{r}, t)$, evolving in the rotating frame at frequency $\omega_p$ according to the driven-dissipative Gross--Pitaevskii equation (GPE) \cite{carusotto2004}
\begin{equation}
i \hbar\partial_t \psi(\mathbf{r},t) = \left( \hbar\Delta - \frac{\hbar^2}{2m} \nabla^2 + g |\psi|^2 - i \frac{\hbar\gamma}{2} \right) \psi + F(\mathbf{r}) ,
\label{eq:GPE1}
\end{equation}
where $g$ is the polariton--polariton interaction constant, the detuning $\Delta = \omega_0^{LP} - \omega_p$ is measured from the bottom $\omega_0^{LP}$ of the lower polariton band of effective mass $m$, and $\gamma$ is the polariton decay rate.

In all the paper we take realistic values of ${\hbar^2}/{m} = 1\,\textrm{meV}\,\mu\textrm{m}^{2}$ and $g = 1\,\textrm{meV}\mu\textrm{m}^{2}$. Dropping for simplicity $\hbar$ factors, we measure all frequencies in units of energy. Notice that, upon a suitable rescaling of $F_0$ and $\psi$, the specific value of $g$ has no relevance for the mean-field dynamics and we can express densities in terms of the (easily experimentally accessible) local blueshift $g|\psi(\mathbf{r})|^2$.


Outside the pump spot, one has $F(\mathbf{r}) = 0$ and it is convenient to also consider a hydrodynamic velocity -- density formalism,
\begin{eqnarray}
\partial_t n &+& \nabla \cdot (n\mathbf{v}) = - \gamma n \\ 
\partial_t \mathbf{v} &+& \nabla \left(  \Delta + \frac{1}{2m} \mathbf{v}^2 + gn - \frac{\hbar}{2m \sqrt{n}} \nabla^2 \sqrt{n} \right)  = 0
\label{eq:en_cons}
\end{eqnarray}
where $n(\mathbf{r},t) = |\psi(\mathbf{r},t)|^2$ and $\mathbf{v}(\mathbf{r},t) = \frac{\hbar}{m} \nabla \arg[\psi(\mathbf{r},t)]$ are the local density and velocity of the fluid.
These equations correspond to density and energy conservation respectively, and, exception made for the decay term $\gamma$, they are identical to the ones used for conservative atomic BECs \cite{pitaevskii2016}.
 
In analogy with the Thomas--Fermi approximation of atomic BECs~\cite{pitaevskii2016}, for sufficiently smooth flow profiles the last ``quantum pressure'' term in eq. (\ref{eq:en_cons}) can be neglected, which gives explicit equations for the steady-state flow profile in one-dimensional configurations
\begin{eqnarray}
\frac{d(gn)}{dx} &=& -  {mv} \frac{dv}{dx}, \label{eq:dgn_dx}\\
\frac{dv}{dx} &=& - \frac{\gamma}{1 - \frac{mv^2}{gn}}
\label{eq:dvdx}
\end{eqnarray}
within a sort of local density approximation (LDA) picture.
From the second equation, we immediatly see that the Thomas--Fermi approximation breaks down in the neighborhood of the trans-sonic point where $mv^2 = gn$ and the derivative of the velocity diverges. On the other hand, when these equations hold, the current profile satisfies the modified continuity equation 
\begin{equation}
{j}({x}) \equiv n {v} = {j}({x_0})\,e^{- \gamma \int_{{x}_0}^{x} \frac{d{x}'}{v({x}')} }
\end{equation}
which includes the effect of the polariton decay.

\section{Outward polariton flow from a localized pump spot}
\label{ssec:1D}

In this section we discuss a first application of the GPE formalism to one-dimensional geometries under a coherent drive with a finite pump spot. In spite of the simplicity of the configuration, interesting features are found for the polariton fluid that propagates away from the pump spot, namely a quick transition to a super-sonic flow and an unexpectedly weak dependence of the flow profile on the pump intensity and wavevector. Polariton wire devices~\cite{wertz2010} appear as ideal candidates where to experimentally investigate this physics in a quantitative way.


\begin{figure*}[]
\includegraphics[width=2.\columnwidth]{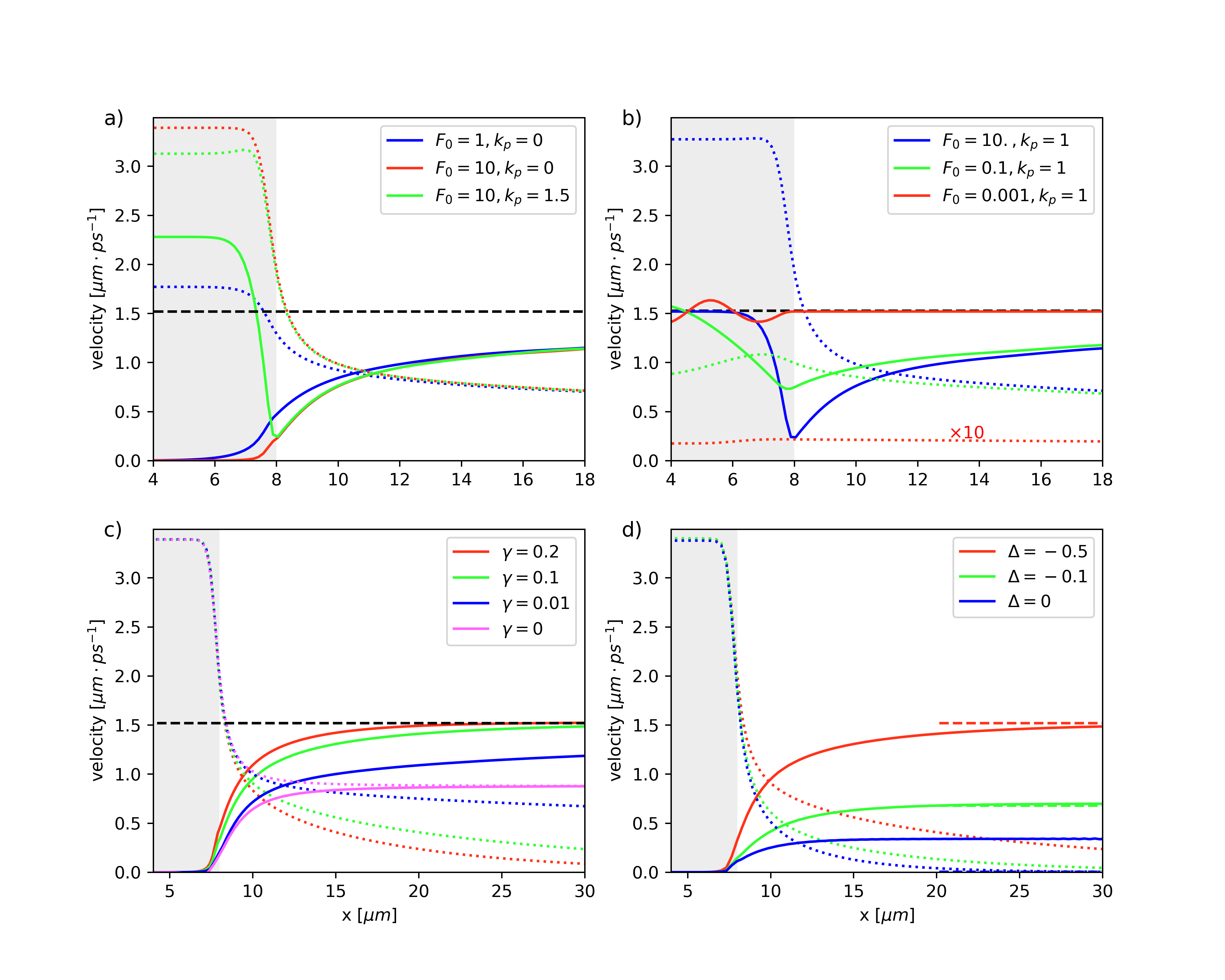}
\caption{Numerical GPE calculations of the one-dimensional polariton flow profile under a spatially localized pump spot. The different panels (a-d) illustrate the dependence on various pump parameters. Continuous lines indicate the flow velocity, while the dotted ones indicate the local speed of sound as defined in (\ref{eq:local_cs}). The horizontal black dashed lines indicates the asymptotic flow velocity $v_\infty=\sqrt{2\,|\Delta|/m}$. 
The grey shaded region indicates the extension of the pump spot, whose profile is given by (\ref{eq:sharp_pump}) with $x_{pe} = 8.0 \mu$m and $\sigma_{pe} = 0.25 \mu$m. If not otherwise specified, $\Delta = -0.5\,$meV, $\gamma = 0.02\,$ meV, $k_p=0$. Units in the legends are meV$ \mu$m$^{-1}$ for the pump amplitude $F_0$ and $\mu$m$^{-1}$ for the pump wavevector $k_p$. In (d), the decay rate is set to $\gamma = 0.1 $ meV. Numerical calculations were performed by means of the 4th order Runge--Kutta method in a simulation box of size $L_x = 1280 \mu$m with a grid of $2^{12}$ points and  periodic boundary conditions.
}
\label{fig:1Dejection}
\end{figure*}

In almost all works so far on superfluidity and vortex nucleation in polariton fluids \cite{amo2011, pigeon2011}, the fluid is quasi--resonantly driven using a finite but large pump spot, and it is more or less implicitly assumed that the flow velocity of the polariton fluid is controlled by the wave vector of the pump $\mathbf{v} = {\hbar} \mathbf{k}_p/{m}$. In this section we show that the physics is significantly more complicate than that.

Far from the pump spot where the polariton density has dropped to a negligible value, the GPE predicts that the asymptotic flow velocity $v_\infty$ is fixed (in modulus) by the (assumed negative) laser detuning according to energy conservation, 
\begin{equation}
\Delta + \frac{m}{2} {v}_{\infty}^2= 0.
\label{eq:v_asym}
\end{equation}
More precisely, in a one-dimensional geometry and on the large-$x$ side of the pump spot, the steady--state wavefunction has the form 
\begin{equation}
\psi({x}) \sim e^{i \frac{m{v}_{\infty} }{\hbar}{x} - \frac{\gamma}{2v_{\infty}}{x}} 
\end{equation}
with 
\begin{equation}
\Delta + \frac{m}{2} v_{\infty}^2  - \frac{1}{8m} \frac{\gamma^2}{ v_{\infty}^2 }=  0. 
\end{equation}
While the last term plays a role for $\Delta \gtrsim 0$, it is fully negligible when the laser frequency is well above the bottom of the polariton band ($-\Delta \gg \gamma$) and the asymptotic speed is correspondingly large.  

Beyond this simple analytical insight on the long-distance behaviour, the overall physics of one-dimensional polariton flows under a spatially finite coherent pump is illustrated in Fig.\ref{fig:1Dejection}. Stationary solutions of the GPE polariton flow emerging from a coherently pumped region delimited by sharp edges are shown for a variety of pump parameters. In all panels, solid lines indicate the spatial profile of the flow velocity ${v}({x})$ defined above, while dotted lines indicate the local speed of sound 
\begin{equation}
c_s({x}) = \sqrt{\frac{g |\psi({x})|^2}{m}}.
\label{eq:local_cs}
\end{equation} 
The pump amplitude $F(x)$ is taken to have a constant wavevector $k_p$ and to be nonzero in the $x<x_{pe}$ region only,
\begin{equation}
F(x) = F_0 e^{ik_px} \left[ 1 -\exp \left\{ -\frac{(x-x_{pe})^2}{2\sigma^2_{pe}}       \right\} \right].
\label{eq:sharp_pump}
\end{equation}
Here, $x_{pe}$ and $\sigma_{pe}$ respectively determine the position and the sharpness of the pump edge.

As a most remarkable feature, panel Fig.\ref{fig:1Dejection}.a shows how the flow profile right outside the pumped region features small variations across a wide range of very different excitation powers and pump wavevectors. This can be clearly seen by looking at the position $ x^*$ of the trans-sonic point such that $c_s(x^*) = v(x^*) $, which does not display major variation for all choices of parameters. Further to this point, the velocity displays a monotonic growth towards the asymptotic value $v_\infty$ that follows by energy conservation (within the local density approximation) the decrease of the polariton density and, thus, of the interaction energy,
\begin{equation}
\frac{m}{2} {v}(x)^2 = -\Delta - g n(x).
\label{eq:energycons}
\end{equation}
On the other hand, the extension of the sub-sonic $c_s(x) > v(x)$ flow region remains quite limited even with a very strong drive and very long lived polaritons and, most importantly, the complex shape of the flow profile in this region is determined by the matching of the long-distance flow mentioned above with the one in the pumped region, whose wavevector and intensity are fixed by the pump via  the usual nonlinear equations for spatially homogeneous configurations~\cite{carusotto2004,carusotto2013}. For instance, depending on the relative value of the pump wavevector $k_p$ and the asymptotic one $k_\infty=m v_\infty/\hbar$, the velocity profile shows a monotonic (red and blue lines) profile if the pump wavevector is smaller than the asymptotic one or a non-monotonic (green) one for larger pump wavevector values: to understand this last behaviour, one needs to remind that at large distances the speed is smaller than the asymptotic speed and grows towards it according to (\ref{eq:energycons}), while in the pumped region it has to match the (larger) pump wavevector $k_p$.
A further mathematical interpretation of this variety of behaviours is found in the denominator of eq. (\ref{eq:dvdx}): imagining to integrate the steady--state with initial condition at large $x$, initially close solutions will strongly deviate from each other as the trans-sonic point $x^*$ is approached.

The specific effect of interactions is illustrated in Fig.\ref{fig:1Dejection}.b, where we fix the pump wavevector to a value close to the asymptotic velocity for the chosen pump frequency and we 
 vary the pump power from the linear regime of very weak pumping towards high intensities well above the bistability loop. In the linear regime (red lines), the density is so low so that $v$ sticks to $v_{\infty}$ as soon as the flow exit the driven spot. The oscillations in the velocity and density profile in the pumped region are due to partial reflection of polaritons at the edge of the pumped region. As it was also experimentally observed in~\cite{nguyen2015}, these oscillations disappear in the strongly nonlinear regime where superfluidity sets in (green and blue lines). In the nonlinear regime, the speed right outside the pump spot is strongly reduced by the higher density according to (\ref{eq:energycons}) and recovers the asymptotic value only at large distances. 

To better understand the peculiar hydrodynamic features in the sub-sonic region, in Fig.\ref{fig:1Dejection}.c we report results for varying decay rate while the pump wavevector and the (relatively strong) intensity are kept constant. No appreciable change is found in the pumped region where the system is far on the upper bistability branch so the density is basically fixed by the interplay of pump intensity and density-dependent detuning~\cite{carusotto2004,carusotto2013}. As expected, the size of the subsonic region up to the trans-sonic point at $x^*$ increases with decreasing $\gamma$, but the dependence is a very slow one and sub-sonic flows of macroscopic size are hardly obtained. 

This statement is corroborated by comparison with the purple curves for the conservative $\gamma = 0$ limit (in the undriven region). In this case absorbing boundary conditions were implemented at large distances by means of a smooth imaginary potential. These curves show that the flow outside the pump spot tends to self--regulate itself to the maximal velocity that can be supported without creating scattering, that is the local speed of sound. Since the non-dissipative condition allows for spatially uniform flows at large distances, the long-distance limit of the speed is now $v_\infty/\sqrt{3}$. Even though the trans-sonic point $x^*$ is now at infinity, the flow is only barely sub-sonic in most of the space. Note in particular how the quick decay of the density right outside the pump spot is not an effect of losses (that are not present here), but originates from hydrodynamics effects dominated by the quantum pressure term.

For non-zero losses $\gamma$, the long-distance decay of the condensate density beyond the trans-sonic point is well captured by the local density picture of Eqs.(\ref{eq:dgn_dx}-\ref{eq:dvdx}) and thus dominated by the decay rate $\gamma$. On the other hand, the transition from the pumped region and the trans-sonic point depends only weakly on $\gamma$ and is qualitatively similar to the behaviour found in the $\gamma=0$ conservative case. This indicates the crucial role of the quantum pressure term that was neglected in Eq.\ref{eq:dvdx} and that is responsible for the fast, almost $\gamma$-independent drop of the density right outside the pump spot.

Finally, in Fig.\ref{fig:1Dejection}.d we show the result of simulations varying the detuning $\Delta$ and keeping fixed the decay rate, the pump wavevector and the inner polariton density (by adjusting the laser strength). For negative enough values of the detuning, the asymptotic velocity is well captured by $v_\infty$ (dashed red and green lines). For $\Delta = 0$, it is instead determined by the decay rate and is equal to $\sqrt{{\gamma}/({2m}})$. Even though both the density and the speed of sound $c_s$ increase with more negative detunings, the behaviour of the fluid  in the outer region is dominated by the even faster growing asymptotic speed, so the size of the sub-sonic region actually decreases with the pump intensity.

The calculations discussed in this section illustrate the dependence of the outward flow on the different pump parameters. From this, one can extract physical insight on the behaviour of the system and anticipate the most favourable regimes to observe a given effect. In particular, if one wants to study superfluidity using a defect located outside the pump spot, it looks beneficial to have $x^*$ as large as possible so to maximize the size of the sub-sonic region. To this purpose, the most favourable regime appears to be at small negative $\Delta$ (and of course small $\gamma$). In the next sections we will see how the physics is somehow richer than that.

\section{Superfluidity and instabilities in two-dimensional flows}
\label{sec:LinGeo}

The features pointed out in the previous section have profound consequences on the hydrodynamics of two-dimensional driven--dissipative fluids generated by a laterally very wide pump (in the $y$ direction) which induces an effectively one--directional flow directed along the $x$ direction. Translational invariance along $y$ is broken by a static obstacle inserted in the flow. 

While the usual behaviour of equilibrium superfluids~\cite{carusotto2013} is observed for obstacles located in the pumped region, the small extension of the sub-sonic region and the quick quantum-pressure-driven density drop discussed in the previous section replace it with a smooth shallow--to--deep soliton crossover for defects located outside the pump. As a result, superfluidity or vortex nucleation are hardly found, even for very small dissipation.
A time-dependent behaviour is recovered when a realistic amount of static disorder is included in the model: the ensuing spatial motion of the phase singularities that occur in the low-density region around the soliton cores is responsible for a localized loss of coherence. Remarkably, this non-stationary behaviour turns out to be facilitated by the presence of a incoherent reservoir.

\subsection{Fragility of the superfluid}
\label{ssec:solcross}

\begin{figure}[htbp]
\includegraphics[width=1.\columnwidth]{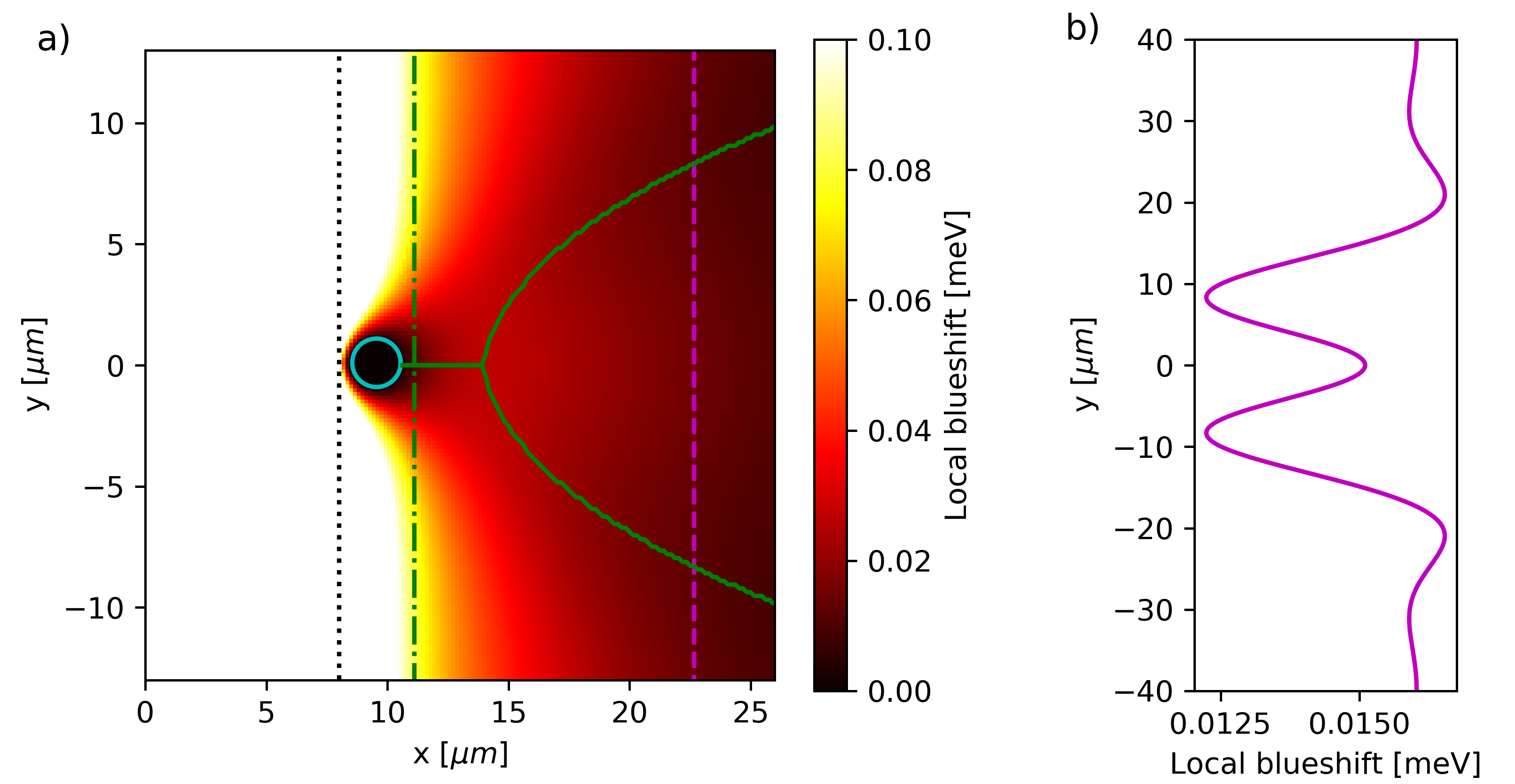}
\includegraphics[width=0.85\columnwidth]{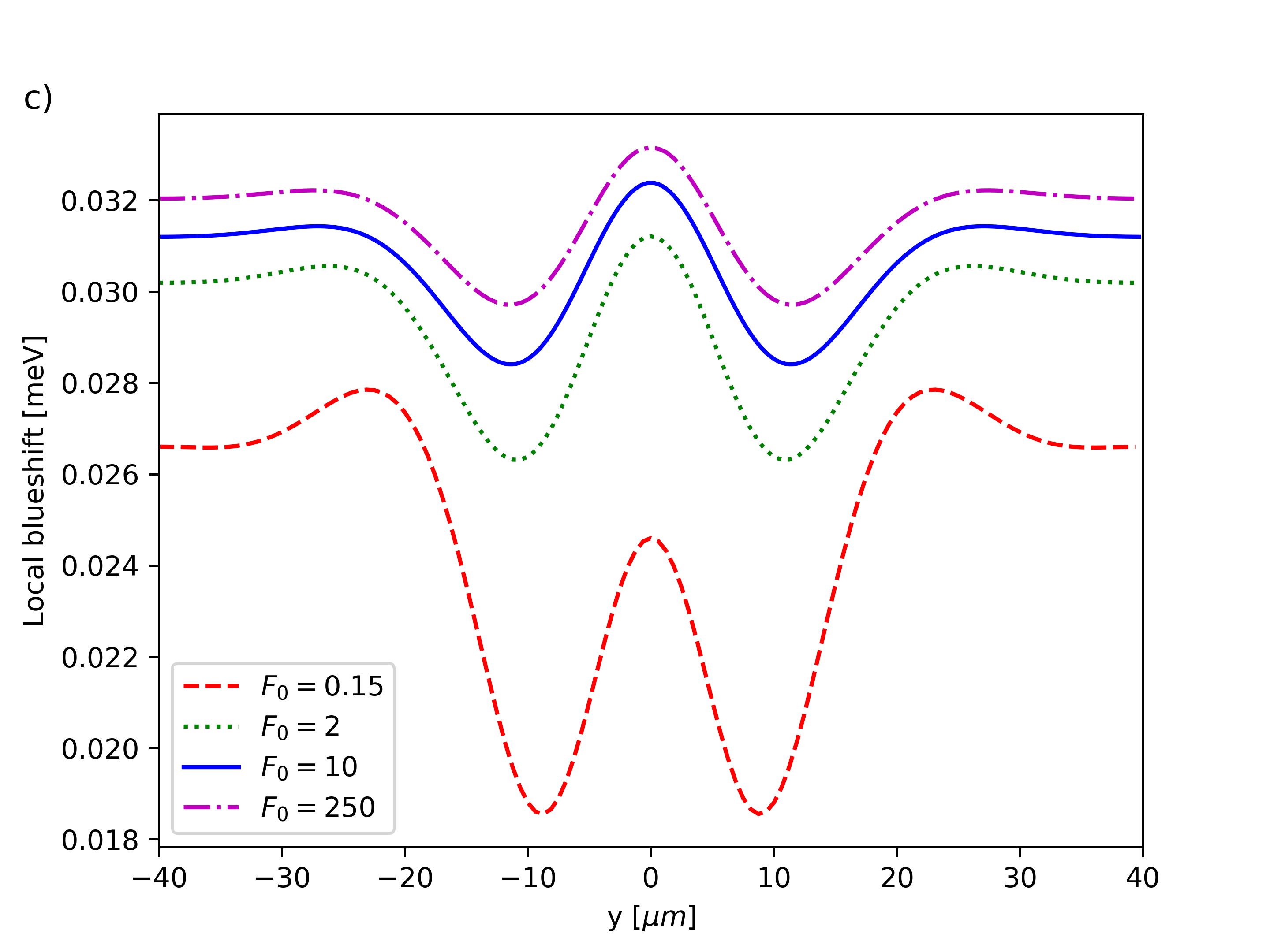}
\includegraphics[width=0.85\columnwidth]{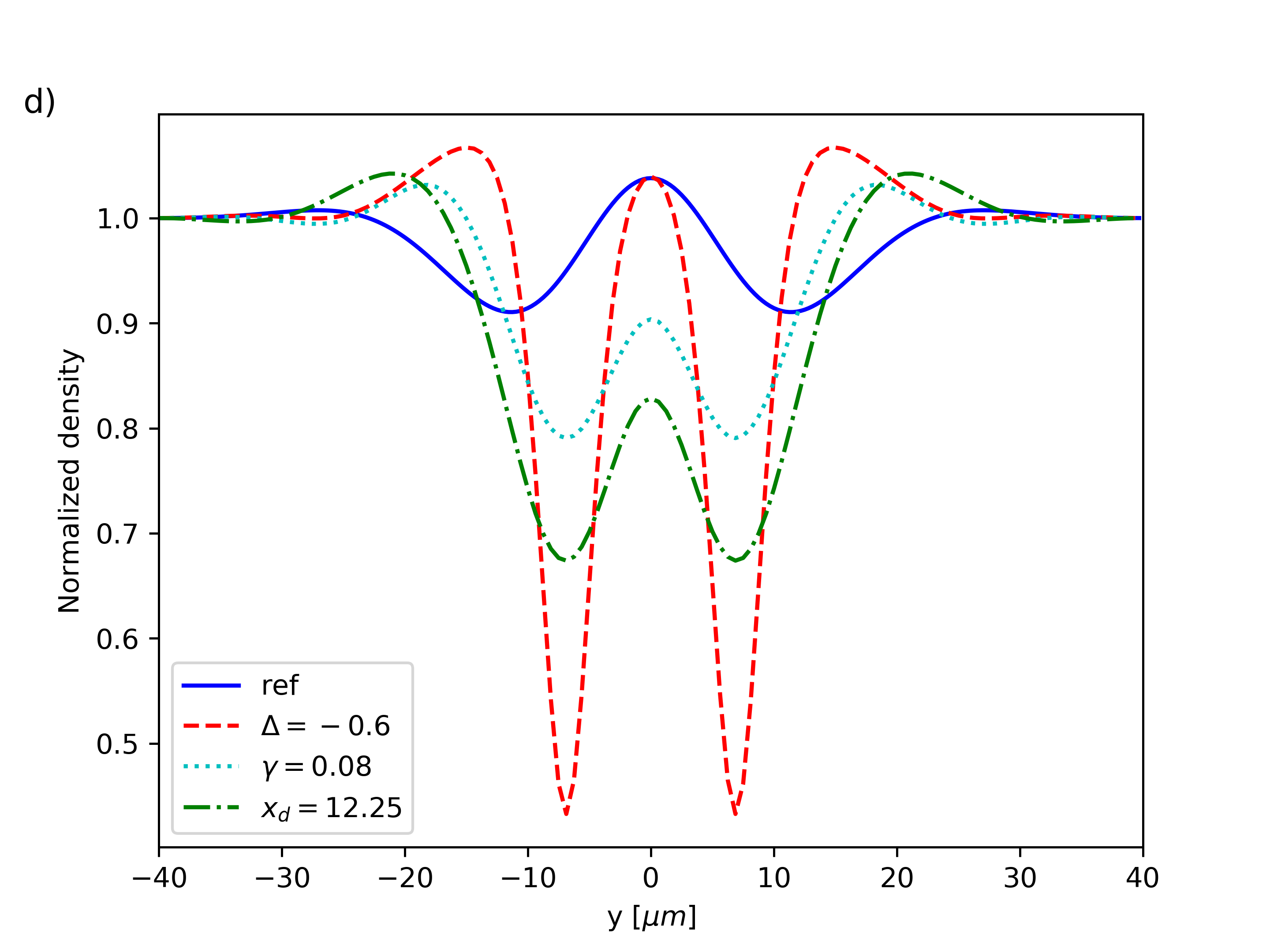}
\caption{Panel (a): Steady-state density profile [in units of the corresponding blue-shift $g n(\mathbf{r})$] of a two-dimensional flow under a coherent pump profile of the form (\ref{eq:sharp_pump}). The vertical black dotted line indicates the edge of the pump spot at $x=x_{pe}$, while the green dash-dotted line indicates the trans-sonic point $x=x^*$ in the absence of the defect. Note that the density is heavily saturated in the pumped region, where the blueshift attains values as high as $0.7$~meV. A cut of the density along the vertical dashed magenta line at $x=23\,\mu$m is shown in panel (b). Parameters: $\Delta = -0.1$ meV, $F_0 = 0.5$ meV$ \mu$m$^{-1}$, $\gamma = 0.04$~meV, $k_p = 1.0 \mu$m$^{-1}$, $x_{pe} = 8.0 \mu$m and $\sigma = 0.25 \mu$m. The defect is of effectively infinite strength $V_d=20$~meV, has radius $r_d=1\,\mu$m and is centered outside the pump spot at $x_d=9.25\,\mu$m.
Panel (c): cuts of the density along the same vertical magenta line at $x=23\,\mu$m for different pump powers. Panel (d) the same cuts when various parameters are varied with respect to the curve in (c) for $F_0 = 10$ meV$ \mu$m$^{-1}$. Given the small polariton decay rate $\gamma_= 0.02$~meV used in (c,d), these simulations had to be performed in a large box of sides $L_x = 640 \mu$m, $L_y = 80 \mu$m with a grid of $2^{11} \times 2^{8}$ points.
}
\label{fig:crossover}
\end{figure}

We consider a pump profile that is a function of $x$ only and does not depend on the lateral coordinate $y$, $F_0=F_0(x)$. The defect is located completely outside of the driven region. The main feature that one can observe in Fig.\ref{fig:crossover} is that a (possibly shallow) V-shaped, soliton-like  density modulation always appears past the defect and extends to the edge of the cloud in the downstream direction. Such a breaking of superfluidity occurs even for weak dissipation, high powers, small defect size, and small distance of this latter from the pumped region. For the case shown in panel (a), the density depletion along the slice marked by the dashed magenta line [shown in panel (b)] is as high as 25\% even though the defect is fully in the region of locally sub-sonic $c_s>v$ flow delimited, in the absence of defect, by the green dash-dotted line. 
Furthermore, in contrast to the conservative case of atomic superfluids~\cite{winiecki2000,neely2010}, no turbulent state with a time-dependent vortex nucleation has been found. Note that this prediction is not in contrast  with the experiment~\cite{nardin2011}, which was performed in a completely different regime where the coherent pump had a pulsed character and the polariton phase was afterwards free to evolve.

Whereas panels (a,b) show the breaking of superfluidity in an experiment with realistic parameters of state-of-the-art samples, panels (c,d) aim at demonstrating that, as soon as dissipation is present, no clean superfluidity is possible for a defect located outside the pump spot. As a first step, in panel (c) we plot sections of the steady-state density for an extremely small decay rate and a series of different values of the pump strength. The cuts are taken far downstream from the defect (along the same line as in the top panel). In agreement with our discussion in Sec.\ref{ssec:1D}, the density does not appear to change much even under a very large (possibly unrealistic) increase of the pump power. Note that the very small value of the decay rate used in this panel requires to perform the simulation in a large box to avoid numerical artifacts~\footnote{The vortical, non--stationary phases reported in \cite{pigeon2011} are likely due to interference of the back--scattered flow by the pumped region through the periodic boundary conditions. Simulations with small decay are in fact very sensitive to scattering effects and we have encountered similar problems when smaller boxes were used. Notice that the box  used here is $16 \times 4$ times larger than the one used in \cite{pigeon2011} with the same numerical grid spacing.}.

We now try to convince the reader that the configuration plotted as a solid blue line in the middle panel is optimal to achieve superfluid-like features for a given decay rate and given the constraints imposed by the presence of dissipation. To this purpose, in panel (d) we present cuts of the density profile at a given $x$ normalized to their large $y$ value for different choices of parameters that differ from the reference one by the variation of a single parameter. Specifically, the blue solid line is the reference, the red dashed line is for a larger detuning ($\Delta = -0.6$ meV instead of $\Delta = -0.1$ meV) for which the flow gets more quickly supersonic, the cyan dotted line is for a larger dissipation ($\gamma=0.08$~meV instead of $\gamma=0.02$ meV) which again reduces the size of the subsonic region, and the green dashed--dotted line is for a defect located farther away from the pump (at $x_d = 12.25 \mu$m instead of $x_d = 9.25 \mu$m), that is closer to the trans-sonic point. 
As discussed in Sec.~\ref{ssec:1D}, the dependence on $k_p$ is a very weak one and thus not worth being illustrated here.

Building on these numerical results, we can trace back the fragility of superfluid flows in the presence of dissipation to the relatively quick decay of the polariton density outside the pump spot studied in Sec.\ref{ssec:1D}. In order to assess whether a defect is able to generate a density perturbation in a spatially inhomogeneous flow, it is in fact not enough to apply the Landau criterion for superfluidity at the defect's location only (which would in any case be questionable in our case, given the strong density gradient in the subsonic region), but one has to look for all points where the flow is supersonic. The quantitative value of the perturbation present at a given point will however depend on the length of sub-sonic flow across which the excitations have to ``tunnel'' before getting ``on-shell'' in the supersonic region. This mechanism is illustrated by the green line in Fig.\ref{fig:crossover}(a) which shows, for each value of $x$, the position along $y$ of the density minima. Since the defect is located upstream of the trans-sonic point, the perturbation is initially evanescent and does not propagate in the lateral direction. Only later, well in the downstream supersonic region, it transforms into the usual, laterally expanding V-shaped structure.

\subsection{Disorder can break stationarity}

The calculations in the previous section have shown how the presence of dissipation induces a remarkable fragility in a superfluid polariton flow. No matter the pump parameters and the spatial position of the defect outside the pump spot, this latter is generally able to induce a significant perturbation in the flow. However, it turned out hard if not just impossible to observe vortex nucleation and the polaritons flow remains generally stationary in time at all points. 

Since experiments~\cite{amo2011} have instead observed a loss of coherence, it is important to assess the importance of other effects on the polariton flow stability. In what follows we will characterize non-stationarity in terms of a first--order coherence function 
\begin{equation}
g^{(1)}(\mathbf{r}) = \frac{|\langle  \psi(\mathbf{r})  \rangle_t|}{\sqrt{     
\langle n(\mathbf{r}) \rangle_t 
  }}, \label{eq:g1}
\end{equation}
which uses the phase of the pump as a reference. As usual, perfect coherence corresponds to having $g^{(2)}=1$.

\begin{figure}[hbtp]
\includegraphics[width=1.0\columnwidth]{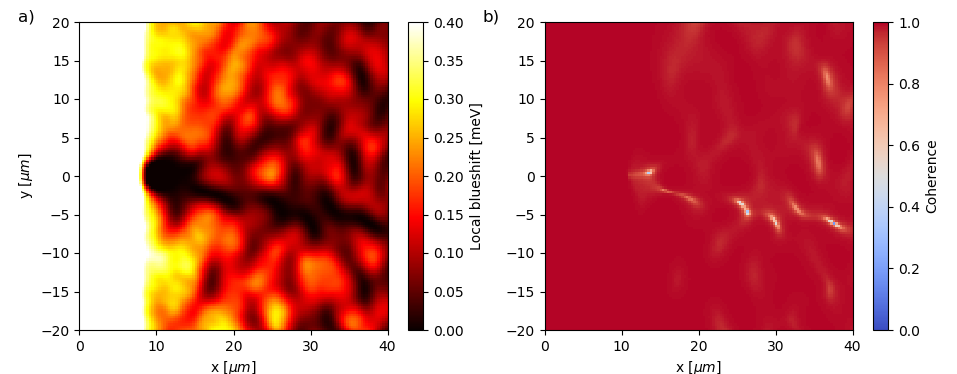}
\caption{(a) Time averaged polariton density and (b) first--order coherence function in the presence of static disorder. This has the Gaussian distribution described in (\ref{eq:gauss_disorder}) with correlation length $\sigma_W = 0.5 \mu$m, and amplitude $W_0 = 0.06$ meV. Pump parameters: $\Delta = -0.4$ meV$, k_p = 0$ and  $F_0 = 0.25$ meV$\,\mu$m$^{-1}$. The strong defect is of radius $r_d = 2 \mu$m and is located at $x_d=10.5 \mu$m.
}
\label{fig:disordered_nucleation}
\end{figure}

Fig.\ref{fig:disordered_nucleation} shows the spatial profiles of the time-averaged density (measured in terms of the local blue-shift $g \langle n(\mathbf{r})\rangle_t$) and of the $g^{(1)}$ coherence function in the presence of a static disorder modeled as a Gaussian-distributed random real potential $W(\mathbf{r})$ with
$
\langle\langle W(\mathbf{r}) \rangle\rangle = 0
$ and
\begin{equation}
\langle\langle W(\mathbf{r}) W(\mathbf{r'}) \rangle\rangle =
W_0^2 e^{- \frac{(\mathbf{r} - \mathbf{r'})}{4\sigma^2_W}},
\label{eq:gauss_disorder}
\end{equation}
where $\langle\langle ... \rangle\rangle$ stands for average over disorder, $\sigma_W$ is the correlation length of the field and $W_0$ its amplitude, chosen of magnitude comparable to the decay rate.

As expected, the  density profile (slighty saturated in the left part of the figure) is perturbed by the presence of disorder. But even more importantly, the fluid often fails to converge to a stationary state and keeps displaying irregular oscillations in time. Further insight in this behaviour is obtained by looking at the time evolution of the phase shown in the Movie M1 included as Supplemental Material~\cite{suppl}: here, one can see how the disorder perturbs the phase fronts and makes them perform small oscillations. This time dependence has the strongest impact in the neighborhood of the vortex-like phase singularities induced by the disorder, in particular in the low density region around the soliton cores. Since the phase displays a singular $2\pi$ variation around the vortex core, any displacement of the vortex has a dramatic effect on $g^{(1)}$. 

\begin{figure*}[hbtp]
\includegraphics[width=2.0\columnwidth]{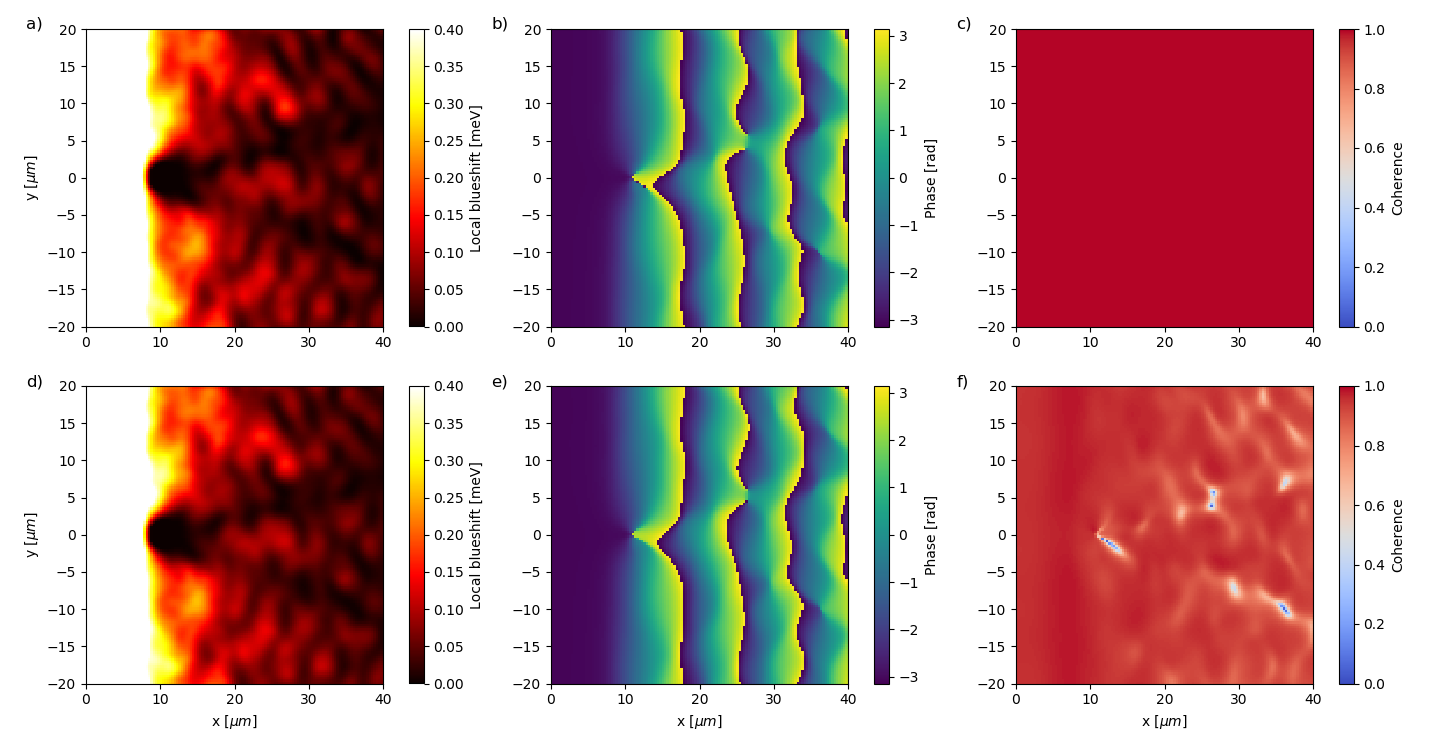}
\caption{Panels (a-c): time-averaged polariton density (a), local phase at the final time (b), and coherence function $g^{(1)}(\mathbf{r})$ for a polariton fluid evolving according to (\ref{eq:GPE1}) in the presence of disorder. Panels (d-f): same quantities for the same parameters in the presence of a incoherent reservoir as in Eqs.(\ref{eq:GPC}-\ref{eq:GPR}). Parameters: $\gamma = 0.04$~meV, $\Delta = -0.4$~meV, $k_p = 0$, $F^{eff}_0 = 0.25$~meV$\,\mu$m$^{-1}$, $W_0 = 0.06$~meV. In panels (d-f), the reservoir has $\gamma_R = 0.01$~meV, $\gamma_{inc} = 0.02$~meV and $g_R=2g$. }
\label{fig:D097}
\end{figure*}

An analogous calculation for a stronger disorder of variance $W_0 = 0.16$ meV instead of $0.06$ meV is displayed in Movie M2~\cite{suppl}. As a general trend, we have found that a stronger disorder is required to break the stationary state for stronger decay rates. For instance, for a doubled decay rate $\gamma = 0.04$ meV close to experimental parameters, no instability is generated for static disorder of amplitude up to around $W_0 = 0.1$ meV. On the other hand, a stationary state with full coherence $g^{(1)} \simeq 1$ over all the box is recovered for a stronger pump. Nonetheless, a density modulation remains clearly visible in the downstream region as soon as the defect is outside the (low-wavevector) pump. 

A stationary state is also found for weak pumps next to the linear regime. Of course, the small speed of sound in this weak interaction regime prevents a superfluid behaviour, so that any defect produces a sizable density modulation.

\subsection{The incoherent reservoir enhances the instability}

Recent experiments~\cite{walker2017,stepanov2018} have given strong evidence that a sizable population of dark excitons is present even under a coherent pump and is responsible for an important fraction to the observed blue shift. In~\cite{stepanov2018} we have theoretically and experimentally highlighted the impact of the incoherent reservoir on the collective excitation modes. Here we address the impact of the reservoir on the dynamical instability of the polariton flow in the presence of disorder and on the consequent loss of coherence. A complementary analysis dealing with the dynamical consequences of the reservoir for superfluidity and uniform flow instabilities is in preparation \cite{amelio2019b}.


As in the previous works, we model the coupled dynamics of the coherent polaritons and of the incoherent reservoir in terms of the paired equations
\begin{eqnarray}
i \partial_t \psi(\mathbf{r},t) &=& \left( \Delta - \frac{1}{2m} \nabla^2 + g |\psi|^2 + g_R n_R +\right. \nonumber \\ &-& \left. i \frac{\gamma}{2} \right) \psi +F(\mathbf{r}) \label{eq:GPC}\\
\partial_t n_R(\mathbf{r},t) &=& - \gamma_R n_R + \gamma_{inc} |\psi|^2,
\label{eq:GPR}
\end{eqnarray}
where $n_R$ is the density of localized dark excitons, $\gamma_R$ is their decay rate and  $\gamma_{inc}$, which contributes to the total polariton decay rate $\gamma = \gamma_c + \gamma_{inc}$, accounts for the conversion of polaritons into reservoir excitons. 

If the reservoir decay $\gamma_R$ were fast, then the reservoir could be adiabatically eliminated as $n_R = g_R  |\psi|^2 {\gamma_{inc}}/{\gamma_R}$ and the presence of the dark exciton would reduce to a mere rescaling of the effective polariton--polariton interaction strength 
\begin{equation}
g_{eff} = g + g_R \frac{\gamma_{inc}}{\gamma_R}.
\label{eq:geff}
\end{equation} 
in the time-dependent equation for the polariton field. However, this fast reservoir condition is not verified in typical systems where the reservoir decay rate $\gamma_R$ is much slower than the polariton decay rate $\gamma$, for instance $\gamma_R \sim 0.015$ meV vs.  $\gamma \sim 0.2$ meV in~\cite{stepanov2018}.
Even if the reservoir can still be eliminated from the equations for the steady-state, we are now going to see how the stability of the steady-state crucially depends on the presence or the absence of the reservoir.


An example of such a behavior is illustrated in Fig.~\ref{fig:D097}. The pump and disorder configuration is chosen in a such way that, in the absence of reservoir, the polariton fluid is dynamically stable. This is illustrated in the panels on the first row. In spite of the clear visibility in the density profiles of soliton-like excitations emerging in the downstream direction from the defect  and of some static vortices pinned by the disorder potential, the coherence function remains basically $g^{(1)}=1$ at all points. 

The panels on the second row show the same quantities in the presence of the reservoir. We have performed a calculation for the two-component model Eqs.(\ref{eq:GPC}-\ref{eq:GPR}) for exactly the same effective parameters: all parameters are equal to the ones of the pure polariton case, including the total effective blue-shift $g_{eff} n = g n + g_R n_R$ which is equal to the polariton blueshift $g n$, but except for the pump intensity that needs rescaling to $F_0 \to F_0\sqrt{{g}/{g_{eff}}}$ to keep the same total blueshift.
As one can see in the figure, introduction of the reservoir drastically changes the long-time behaviour of the system. The soliton-like dark stripes are very similar, but the vortices are now oscillating in time. This results in a suppressed coherence function $g^{(1)}$, in particular along the soliton lines.




\section{Further considerations}
\label{sec:CircularSpot}

In this last Section, we complete our discussion by considering additional configurations whose interesting features may be of importance to understand experimental observations.

\subsection{Long-tailed pump spot}
\label{ssec:fringes}

 \begin{figure}[hbtp]
\includegraphics[width=1.0\columnwidth]{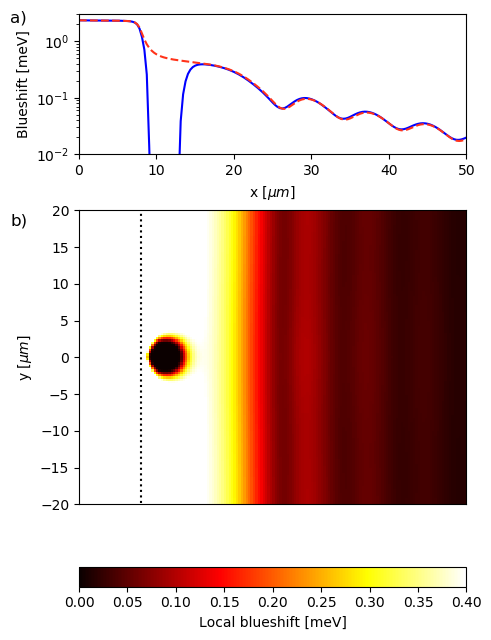}
\caption{Main (b) panel: density profile under a long-tailed coherent pump as described in Eq.(\ref{eq:tanpump}) with $x_{pe} = 8 \mu$m (indicated by the vertical black dotted line) and $\zeta = 0.25 \mu$m. The defect is centered at $x_d = 11 \mu$m and has a radius $r_d= 2 \mu$m. Parameters $F_0 = 3.0$ meV$ \mu$m$^{-1}$, $\gamma = 0.05$~meV, $\Delta = -0.4$~meV and $k_p=0$.
Upper (a) panel: cuts of the local blueshift along the $y=0$ (solid blue) and $y=20 \mu$m (red dashed) lines.
}
\label{fig:fringes}
\end{figure}

We will start from the case of a pump profile whose intensity is mostly concentrated in a small region of space, but also presents a long tail extending outside the main spot. Unless specific efforts are made to restrict the light intensity to a finite spot, in any experiment diffraction is in fact likely to be responsible for a weak light intensity to be present in the whole space.

This physics is illustrated in Fig.\ref{fig:fringes}. We focus on a model configuration where the pump spot is assumed to be uniform along $y$ while, along $x$ it has a sharp jump around $x_{pe}$ and a long tail that slowly decays towards infinity,
\begin{equation}
F(x) = \frac{F_0}{\pi} \left[ \frac{\pi}{2} + \textrm{atan} \left( \frac{x_{pe} - x}{\zeta} \right) \right].
\label{eq:tanpump}
\end{equation}
In contrast to pump with a finite spatial support, the fact that the pump wavevector locks the velocity at all points facilitates the observation of a robust superfluidity. This robustness is illustrated in the figure by the absence of a density modulation around the defect, even though this latter is located well outside the high-intensity part of the pump spot delimited by the vertical black dotted line.

As another interesting feature, we notice clear vertical fringes in the right part of the figure. Such fringes were visible in the experiment~\cite{stepanov2018} but do not appear when the pump has a finite support or has a fast decaying, e.g. Gaussian, tail. Rather than shock waves, a natural interpretation for these fringes is in terms of the interference between the tail of the coherent pump with the polaritons that are simultaneously ejected from the high-density region according to the outward flow mechanism discussed in Sec.\ref{ssec:1D}. An evidence for this mechanism is provided by the approximate $\sqrt{-\Delta}$ dependence of the fringe wavevector on the detuning $\Delta$ for vanishing pump wavevector $k_p=0$ (not shown). Note that a related but different mechanism for controlling superfluid and turbulent behaviours was investigated in~\cite{pigeon2017} using a support field on the downstream side.

\subsection{Circular spot geometry}

\begin{figure}[hbtp]
\includegraphics[width=0.8\columnwidth]{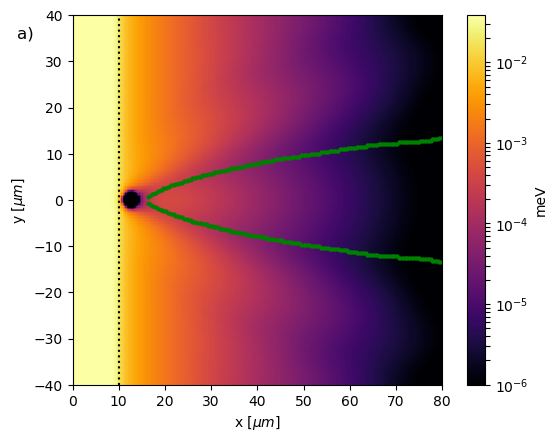}
\includegraphics[width=0.8\columnwidth]{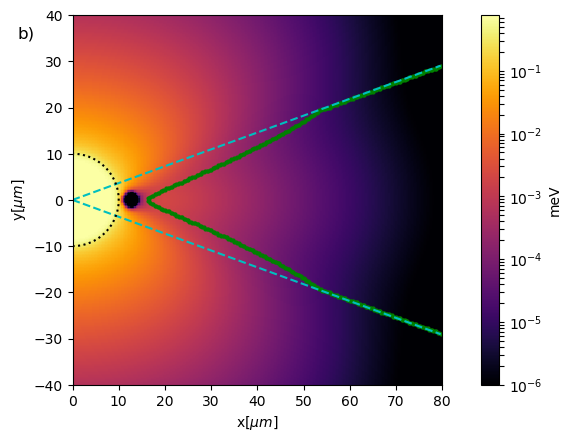}
\caption{Polariton density profile under the linear pump spot of Eq.(\ref{eq:sharp_pump})  [upper panel (a)] and under the circular pump spot of Eq.(\ref{eq:sharp_circular}) [lower panel (b)]. Black dots indicate the edge of the pumped region, while the green points indicate for each $x$ the position of the density minima. Parameters: $\Delta = -0.2$~meV, $\gamma = 0.08$~meV, $F_0 = 0.5$~meV$\,\mu$m$^{-1}$,   $k_p = 0$,
$r_{pe} = 10\,\mu$m, $\sigma_{pe}=0.25\,\mu$m and $x_d = 12.5\,\mu$m. 
}
\label{fig:angle_solitons}
\end{figure}

We conclude our discussion with some remarks on the experimentally relevant case of a finite circular laser spot,
\begin{equation}
F(\mathbf{r}) = F_0 e^{ik_p r} \left[ 1 -\exp \left\{ -\frac{(r-r_{pe})^2}{2\sigma^2_{pe}}       \right\} \right]
\label{eq:sharp_circular}
\end{equation}
for $r<r_{pe}$ and $0$ for $r>r_{pe}$.
In this geometry, illustrated in Fig.\ref{fig:angle_solitons}, reaching the superfluid regime is even more difficult than in the linear geometry of Fig. \ref{fig:crossover}. This is a direct consequence of the geometrical spreading of an outgoing circular wave, which results in a reinforced decay of the density along the radial direction.

Another interesting consequence of the radial flow are the different shapes of the solitonic dark lines in the absence of disorder in the two cases of a linear or a circular pump described by Eqs.(\ref{eq:sharp_pump}) and (\ref{eq:sharp_circular}) and illustrated in the upper and lower panels, respectively. Here, the black dots indicate the geometrical boundary of the pump spot, while the green points mark for each $x$ the position of the density minimum.

While the soliton pair starts from the defect with a similar aperture in the two cases, their shape becomes markedly different further downstream. For a linear pump spot (upper panel), the rapid growth of the flow speed over the speed of sound leads to soliton lines that asymptotically approach the $x$ direction. For a circular pump spot (lower panel), the soliton angle eventually approaches the radial direction of the flow (cyan dashed lines) and has the quite rectilinear shape observed in the experiments~\cite{amo2011}.

\section{Discussion and Conclusions}
\label{sec:conclu}

In this article we have studied the novel features displayed by polariton superfluidity under a spatially localized coherent pump. In contrast to the case of a spatially homogeneous pump considered in~\cite{carusotto2004}, the driven-dissipative condition and the quantum pressure term beyond the local density approximation are responsible for a quick drop of the polariton density right outside the pump spot. This effect reduces the spatial extension of the subsonic flow region and suppresses its superfluidity features. As a most striking feature, a sizable soliton-like density modulation appears downstream of a static defect located outside the pump spot but still in the subsonic region. The impact of a static disorder and of a incoherent reservoir on the dynamical stability of the flow and on the development of turbulent behaviours is highlighted. 

Our results highlight the importance of a new generation of experiments going beyond the pioneering experimental studies of polariton superfluidity~\cite{amo2009,amo2011,sanvitto2011,nardin2011} and exploring polariton hydrodynamic in a quantitative way. These new experiments are needed to firmly assess the mechanisms at play in realistic microcavity devices and possibly to isolate new phenomena not yet included in present non-equilibrium Gross-Pitaevskii equation models, due e.g. to the exciton reservoir and its spatial dynamics~\cite{caputo2017}, interaction with the static disorder, and polariton scattering on phonons~\cite{Savenko2013}. This analysis will be of paramount importance in view of using polariton fluids as a platform for quantum simulation of the physics of quantum fluctuations beyond mean-field such as analog Hawking radiation~\cite{gerace2012,nguyen2015}.

\section*{Acknowledgements}

We are grateful to Simon Pigeon for useful discussions. We acknowledge financial support from the European Union FET-Open grant ``MIR-BOSE'' (n. 737017), from the H2020-FETFLAG-2018-2020 project ``PhoQuS'' (n.820392), and from the Provincia Autonoma di Trento. All numerical calculations were performed using the Julia Programming Language \cite{julia}.

\bibliography{bibliography}

\end{document}